%% file: main.tex
\PassOptionsToPackage{dvipsnames,table}{xcolor}

\documentclass[conference,dvipsnames,table]{IEEEtran}
\input{packages}

\begin{document}

\title{Jupyter Notebook Attacks Taxonomy: Ransomware, Data Exfiltration, and Security Misconfiguration\\ }

\author{\IEEEauthorblockN{Phuong Cao\thanks{*Corresponding author: Phuong Cao; Data: \href{https://pmcao.github.io/pqc}{https://pmcao.github.io}} \orcidlink{0000-0001-6028-0583}$^{1,2,*}$
\\}
\IEEEauthorblockA{
\small{
$^1$National Center for Supercomputing Applications,
$^3$Coordinated Science Laboratory,
$^3$University of Illinois at Urbana-Champaign
}}
}

\maketitle

\input{00_abstract}

\begin{figure*}
    \includegraphics[width=0.95\textwidth]{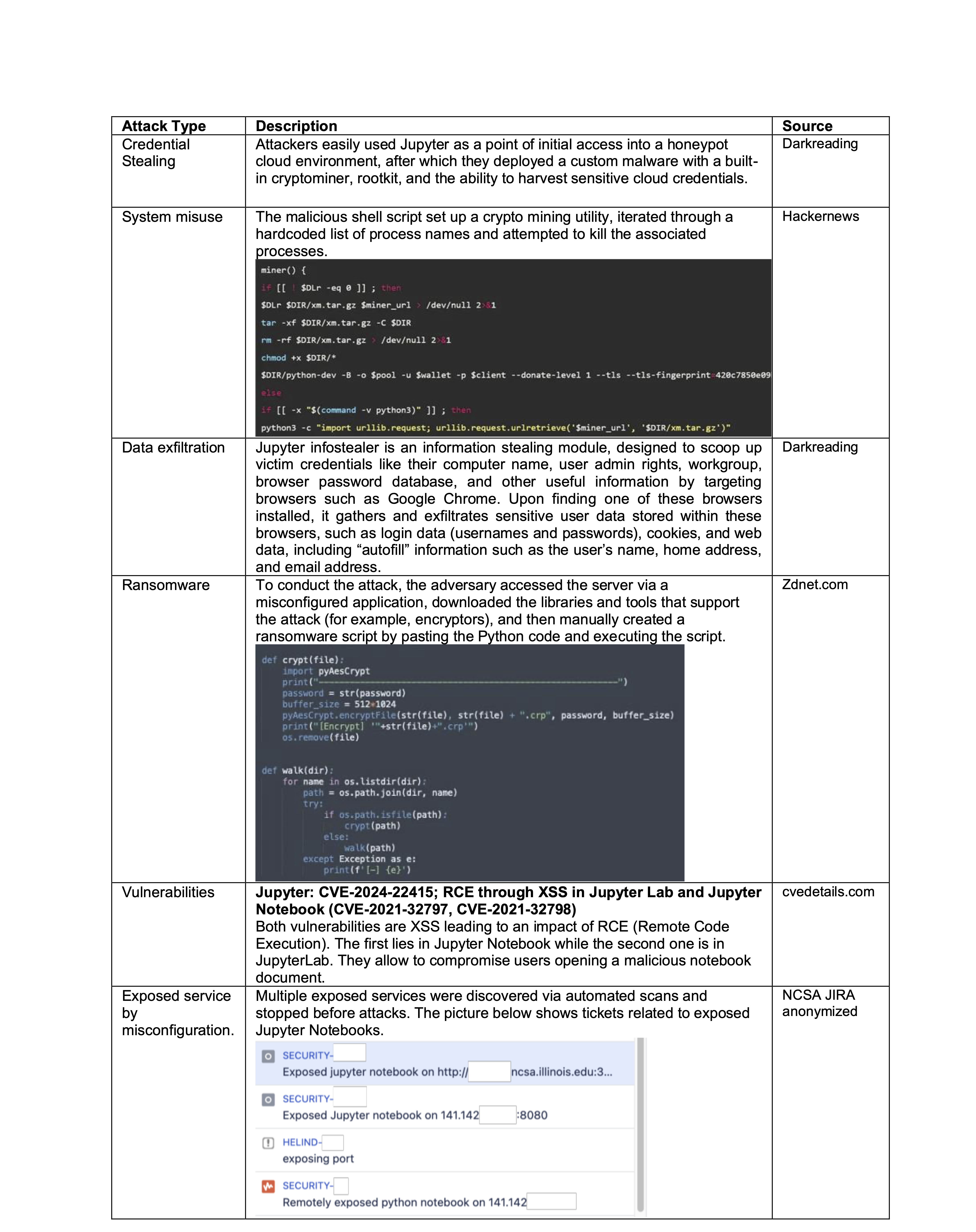}
\caption{Taxonomy of threat models following TrustedCI's Open Science Cyber Risk Profile.}
    \label{fig:taxonomy}
\end{figure*}

\input{01_introduction}

\input{03_background}
\input{04_model}
\input{06_lessons}
\input{07_related_work}

\input{08_conclusion}
\input{99_ack}

{
    \bibliographystyle{IEEEtran}
    \bibliography{references}
}
\end{document}

%% file: packages.tex
\usepackage{xcolor}
\definecolor{myred}{RGB}{255,0,0}
\definecolor{myblue}{RGB}{0,0,255}

\usepackage{pdfpages}
\usepackage{orcidlink}

\usepackage{amsmath,amsfonts,amssymb}
\usepackage{mathtools}
\usepackage{amsthm}
\usepackage{alltt, xspace, times, epsfig}
\usepackage{gensymb,xfrac,array,booktabs,bm}
\usepackage[linesnumbered,lined,ruled]{algorithm2e}
\usepackage{adjustbox}
\usepackage{booktabs}
\usepackage{soul}  

\usepackage{mathtools}
\usepackage{amsmath}

\usepackage[utf8]{inputenc}
\usepackage[english]{babel}

\usepackage{listings}

\usepackage{multirow, multicol, booktabs, tabulary, tabu, longtable, array, varwidth}
\usepackage{placeins, lipsum}
\usepackage[flushleft]{threeparttable}

\usepackage{soul}

\usepackage{tabularx}

\usepackage{amssymb}

\usepackage[labelfont=bf]{caption}
\usepackage{cite}

\usepackage[dvipsnames, table]{xcolor}
\hypersetup{
    linkcolor  = blue!85!black,
    citecolor  = blue!85!black,
    urlcolor   = blue!85!black,
    colorlinks = true,
    breaklinks = true
}

\usepackage{enumerate}

\usepackage[inline]{enumitem}

\usepackage{amsmath, amsfonts, amssymb, amsthm, nicefrac}
\usepackage{mathptmx}  
\usepackage[mathcal]{eucal}

\theoremstyle{definition}

\theoremstyle{comment}

\usepackage[]{cleveref} 
\crefname{appsec}{Appendix}{Appendices}
\crefformat{section}{\S#2#1#3}
\crefformat{subsection}{\S#2#1#3}
\crefformat{subsubsection}{\S#2#1#3}
\crefformat{equation}{(#2#1#3)}
\Crefformat{equation}{Equation~(#2#1#3)}
\crefrangeformat{equation}{(#3#1#4--#5#2#6)}
\Crefrangeformat{equation}{(#3#1#4--#5#2#6)}
\crefmultiformat{equation}{(#2#1#3)}{ and~(#2#1#3)}{, (#2#1#3)}{ and~(#2#1#3)}
\crefformat{figure}{Fig.~#2#1#3}
\Crefformat{figure}{Fig.~#2#1#3}
\crefmultiformat{figure}{Figs.~#2#1#3}{ and~#2#1#3}{, #2#1#3}{ and~#2#1#3}
\crefformat{algorithm}{Algorithm~#2#1#3}
\Crefformat{algorithm}{Algorithm~#2#1#3}
\crefmultiformat{algorithm}{Algorithm~#2#1#3}{ and~#2#1#3}{, #2#1#3}{ and~#2#1#3}
\crefformat{table}{Table~#2#1#3}
\Crefformat{table}{Table~#2#1#3}
\crefrangeformat{table}{Tables~#3#1#4--#5#2#6}
\Crefrangeformat{table}{Tables~#3#1#4--#5#2#6}
\crefmultiformat{table}{Tables~#2#1#3}{ and~#2#1#3}{, #2#1#3}{ and~#2#1#3}

\usepackage{textcomp}
\usepackage[shortcuts,acronym]{glossaries}
\usepackage{soul, footnote, xargs}
\usepackage[colorinlistoftodos,prependcaption,textsize=tiny]{todonotes}
\usepackage{pifont}
\usepackage{balance}

\newcommand\thefont{\expandafter\string\the\font}

\usepackage{tcolorbox}

\usepackage{graphicx}
\usepackage[tight,footnotesize]{subfigure}
\usepackage{tikz, epstopdf, stfloats, bbding, capt-of}
\usepackage{algorithmic}
\usepackage[ruled, linesnumbered]{algorithm2e}

\SetCommentSty{mycommfont}
\graphicspath{{figs/}}
\renewcommand{\footnotesize}{\scriptsize}

\usepackage{wasysym}

\definecolor{colortx}{HTML}{FABD2F}

\definecolor{textgray}{gray}{0.4}
\usepackage[%
    font={small},
    labelfont=bf,
]{caption}

\usepackage{xstring}

\newcommand{\para}[1]{
\vspace{2px}
\noindent{\bf \IfEndWith{#1}{.}{#1}{#1.}}
}

\usepackage{breqn}

\usepackage{amsmath}

%% file: 00_abstract.tex
\begin{abstract}
Open-science collaboration using Jupyter Notebooks may expose expensively trained AI models, high-performance computing resources, and training data to security vulnerabilities, such as unauthorized access, accidental deletion, or misuse. The ubiquitous deployments of Jupyter Notebooks ($\approx$ 11 million public notebooks on Github have transformed collaborative scientific computing by enabling reproducible research. Jupyter is the main HPC's science gateway interface between AI researchers and supercomputers at academic institutions, such as the National Center for Supercomputing Applications (NCSA), national labs, and the industry. An impactful attack targeting Jupyter could disrupt scientific missions and business operations.

This paper describes the network-based attack taxonomy of Jupyter Notebooks, such as ransomware, data exfiltration, security misconfiguration, and resource abuse for cryptocurrency mining. The open nature of Jupyter (direct data access, arbitrary code execution in multiple programming languages kernels) and its vast attack interface (terminal, file browser, untrusted cells) also attract attacks attempting to misuse supercomputing resources and steal state-of-the-art research artifacts. Jupyter uses encrypted datagrams of rapidly evolving WebSocket protocols that challenge even the most state-of-the-art network observability tools, such as Zeek. 

We envisage even more sophisticated AI-driven attacks can be adapted to target Jupyter, where defenders have limited visibility. In addition, Jupyter's cryptographic design should be adapted to resist emerging quantum threats. On balance, this is the first paper to systematically describe the threat model against Jupyter Notebooks and lay out the design of auditing Jupyter to have better visibility against such attacks.

\end{abstract}

%% file: 01_introduction.tex
\section{Introduction} 
Open-science collaboration using Jupyter notebooks may expose expensively trained AI models, high-performance computing resources, and training data to security vulnerabilities, such as unauthorized access, accidental deletion, or misuse. The ubiquitous deployments of Jupyter Notebooks ($\approx$ 11 million public notebooks on Github~\cite{GitHubpa45:online}) have transformed collaborative scientific computing by enabling reproducible research~\cite{DataScie35:online}. Major facilities such as DoE Argonne AI Testbed, the National
Center for Supercomputing Applications (NCSA) Delta GPU and supercomputers such as NERSC's Perlmutter, SDSC's rely on Jupyter~\cite{JupyterH23:online} to provide HPC's science gateway interface for AI researchers. 

This paper highlights the importance of auditing Jupyter Notebooks deployment, which is critical to the operational security of flagship NSF cyberinfrastructure, for example, supercomputers such as Expanse (SDSC), Perlmutter (NERSC), Leadership-Class Computing Facility (LCCF) such as Frontier (ORNL), and related projects such as ACCESS at NCSA/UIUC, Open OnDemand at OSC all rely on Jupyter to advance collaborative and reproducible science research. 

\textbf{Background.} Jupyter notebooks represent code, results, and notes of different scientific applications using JavaScript Object Notation (JSON) documents, which are language-independent, structured text-based file formats. A JSON string represents each cell in a Jupyter Notebook. Notebooks can be processed by any programming language through kernels (Python, R, or Julia). Jupyter uses encrypted datagrams of rapidly evolving WebSocket protocols~\cite{websocke22:online} that challenge even the most state-of-the-art network observability tools, such as Zeek. 

\textbf{Threat model.} A taxonomy of network-based attacks targeting a Jupyter Notebook is described. The open nature of Jupyter (direct data access, arbitrary code execution in multiple programming languages kernels), tight integration with single sign-on~\cite{9356415,chen2017svauth}, and its vast attack interface (terminal, file browser, untrusted cells) also attracts attacks attempting to misuse supercomputing resources and steal state-of-the-art research artifacts (e.g., CVE-2024-22415 and ~\cite{NVDCVE2032:online,NVDCVE2087:online}). These threats, such as ransomware, data exfiltration, security misconfiguration, and resource abuse for cryptocurrency mining, are frequently seen in traditional high-performance computing infrastructure~\cite{cao2014personalized,cao2014preemptive,cao2019preempting} but have evolved to specifically target Jupyter. 

We envisage even more sophisticated AI-driven attacks ~\cite{10224946} can be adapted to target Jupyter, where defenders have limited visibility. For example, resource constraints and scalability challenges are the limiting factors as network traffic keeps increasing, and security monitors may add unsustainable performance overhead to scientific computing. Another example is an evasion attack against the integrity of security monitors. Attackers may employ Denial of Service attacks and inferring detection rules using adversarial machine learning. 

\textbf{Contributions:} 

\begin{itemize}
    \item A taxonomy of existing threats that have targeted Jupyter Notebooks deployments, particularly in supercomputing/HPC environments.
    \item A detailed threat model targets Jupyter Notebooks following TrustedCI's Open Science Risk Profile (OSRP). 
    \item Lessons learned in auditing Jupyter Notebooks's security architecture.
    \item Discussions of hardened security design and futuristic threats targeting Jupyter Notebooks.
\end{itemize}

\newpage
\textbf{Putting this paper in perspective.} To the best of our knowledge, this is the first systematic study highlighting the network security issues of Jupyter Notebooks, focusing on supercomputing/HPC environments. Carefully auditing such threats is critical for securing data-intensive scientific workloads.

%% file: 03_background.tex
\begin{figure*}[t!]
    \includegraphics[width=0.9\textwidth]{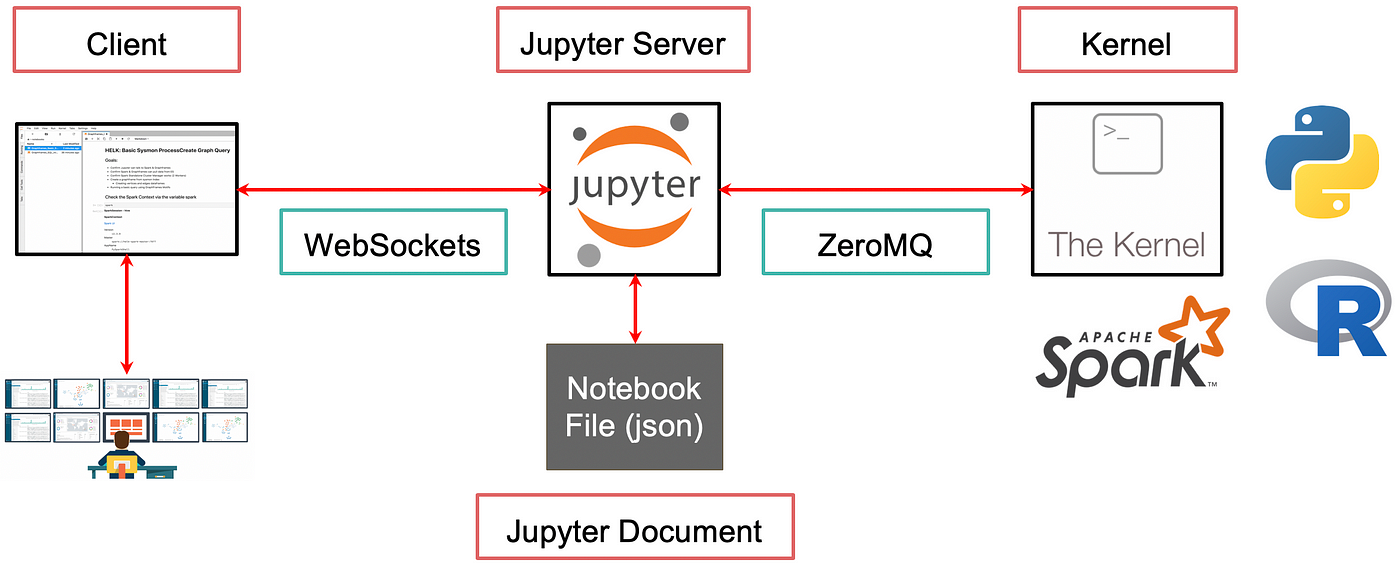}
\caption{Illustration of a Jupyter Notebook's communication flow, based on Jupyter's internal documentation~\cite{Messagin41:online}.}
    \label{fig:background}
\end{figure*}

\section{Background: \\ kernel communication protocols in Jupyter}
This section describes the communication protocols of front-end applications and back-end kernels in Jupyter Notebooks, allowing us to understand different attack entry points.

Jupyter notebooks represent code, results, and notes of different scientific applications using JavaScript Object Notation (JSON) documents, which are language-independent, structured text-based file formats. A JSON string represents each cell in a Jupyter Notebook. Notebooks can be processed by any programming language through kernels (Python, Julia, R, etc.) and converted to other formats such as Markdown, HTML, LaTeX/PDF, and others.

Jupyter implements a two-process model (~\cref{fig:background}) with a kernel and a client for the internal processing of a particular application. In addition, external users communicate with the application, e.g., by writing codes or issuing commands, using secure transport such as Hypertext Transfer Protocol Secure (HTTPS). The specific communication protocol is described in Jupyter's documentation~\cite{Messagin41:online} as follows.

\textit{Internal communication in Jupyter following the Read-Evaluate-Print Loop (REPL) model.} The client takes the user's code specified in a cell, sends it to the corresponding kernel to execute, and returns the result to the client for display. The client can be implemented with a Qt widget or a browser to serve a web-based interface, Jupyter Notebook. This architecture decouples the linear document containing the notebook and the underlying kernel.

\textit{External communication with users.} All communication procedures between Jupyter's internal processes and external users are implemented based on ZeroMQ (or ZMQ) messaging protocol and transported using WebSocket, a TCP-based protocol intended for real-time communication supported by web browsers. For example, Jupyter listens on several ports \texttt{shell\_port, iopub\_port, control\_port, hb\_port} using TCP transport with HMAC-SHA256 signature.

%% file: 04_model.tex
\section{Threat Model}
\label{sec:model}
This section analyzes the vulnerabilities of Jupyter Notebooks, outlining potential attack vectors, adversary techniques, and defensive gaps.

\begin{figure*}[ht!]
    \includegraphics[width=0.95\textwidth]{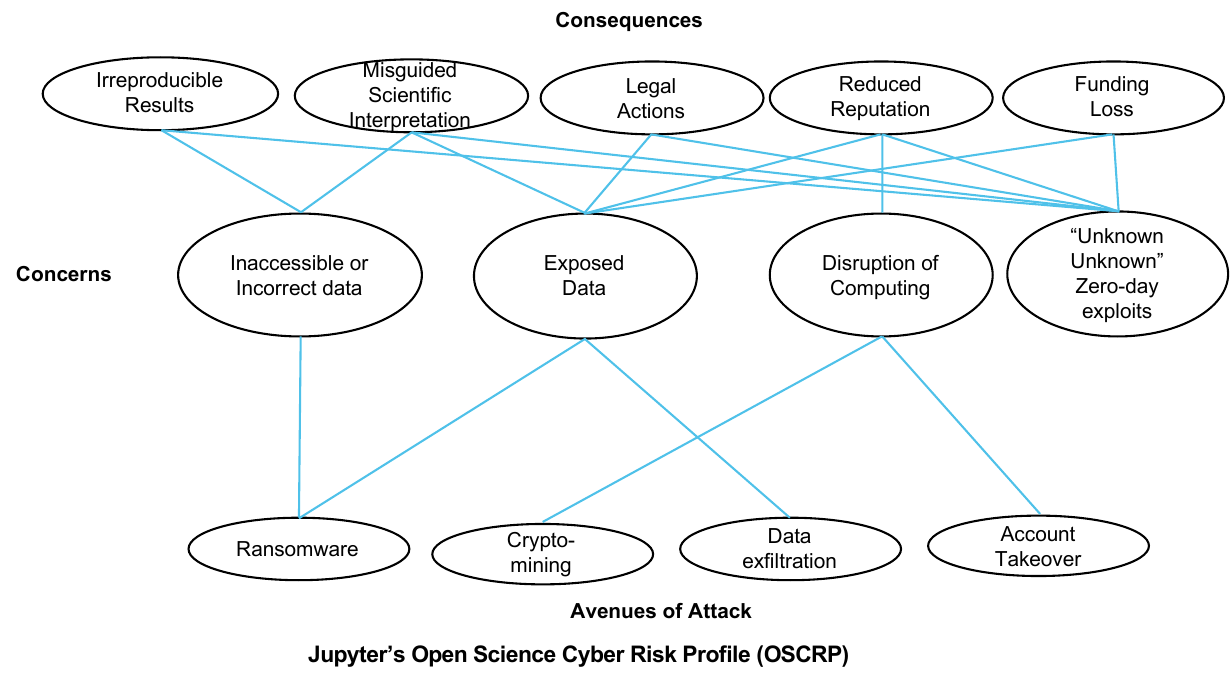}
\caption{Taxonomy of threat models following TrustedCI's Open Science Cyber Risk Profile.}
    \label{fig:oscrp}
\end{figure*}

Following TrustedCI's Open Science Cyber Risk Profile (OSRP~\cite{8055662}) and the TrustedCI Framework, we have identified key data assets, concerns about facilities \& human, and software risks in Jupyter -- which constitute our initial threat model as described in ~\cref{fig:oscrp}. The avenues of attacks include ransomware, crypto mining, and data exfiltration (Table 1), which raise concerns about disruption of computing and inaccessible/incorrect data. 

The threats to science are irreproducible results and misguided scientific interpretation. The consequences to facilities and humans are legal actions, reduced reputation, and funding loss.

In addition, ~\cref{fig:taxonomy} shows i) a taxonomy of Jupyter attacks in the wild that we have collected and ii) internal Jupyter security issues regarding science assets that NCSA and its partners have identified and studied. Note that this is an incomplete and preliminary list of attacks in the scope of this workshop paper.


%% file: 06_lessons.tex

\section{Lessons learned and Discussions}

\subsection{Lessons learned}
This section describes the lessons learned regarding the risk factors of auditing and monitoring Jupyter Notebooks. 

\begin{itemize}
    \item \textit{Resource constraints and scalability challenges.} Network traffic will keep increasing, and a security auditor may add unsustainable performance overhead to scientific computing. To address this, one must  
    harness the power of supercomputers and security domain experts to alleviate this performance overhead.
    
    \item \textit{Evasion attacks against the integrity of security monitors.} Attackers may employ techniques such as low and slow DoS and inferring detection rules using adversarial machine learning. This is a cat-and-mouse game. Defenders aim to stay ahead of attackers by deploying Jupyter Notebook monitors early at the network edges, for example, on a set of honeypots~\cite{cao2019caudit,10333917}, to catch the latest signatures of attacks in the wild -- before they reach the actual Jupyter Notebooks instances deployed in supercomputers. 
    
\end{itemize}

\subsection{Discussions}
This section describes the need for auditing tools and an open-source dataset to innovate Jupyter Notebooks security capabilities. 

\textbf{Tools:}

\begin{itemize}
    \item \textit{Jupyter network monitoring tool:} A network monitoring system is needed to identify malicious users masquerading as legitimate ones in Jupyter notebooks. This tool must collect the honeypot data and threat intelligence sharing infrastructure learned from the edge for effective detection.
    
    \item \textit{Jupyter kernel auditing tool:} An embedded tracing tool must be embedded in Jupyter kernel (starting with Python kernel) to enable extensive logging of user commands.

\end{itemize}

Both of these tools should be integrated into the main branch of the Jupyter project.

\textbf{Dataset:}
\begin{itemize}
    \item \textit{Jupyter Security \& Resiliency Data Set:} There is a clear need for an open-source dataset of Jupyter-related logs in the scientific data workloads. Although NCSA can retain longitudinal data, log anonymization and privacy-preserving sharing need to be studied.
    \end{itemize}
    
We anticipate new threats with evolving computer architectures, such as quantum computing and AI-driven attacks. First, the cryptographic infrastructure of Jupyter Notebooks, being deployed in high-assurance environments, must resist quantum computers. Two immediate threats are posed: harvest now, decrypt later, and digital signature spoofing~\cite{sowa2024postquantumcryptographypqcnetwork,giusto2024dependableclassicalquantumcomputersystems}. Second, attacks driven by generative AI tools will automate our listed threats above and increase the volume of attacks, further challenge the security monitoring system~\cite{cao2024securitytestbedpreemptingattacks}.

%% file: 07_related_work.tex
\section{Related Work}
No existing security monitoring architecture has been specifically developed and tailored to Jupyter, despite the above-growing security threats. Existing Jupyter's code base only has basic security features such as password-based and token-based authentication~\cite{Issues·j56:online}. Federated authentication, such as OpenID Connect (OIDC), is provided by third-party developers with minimal guarantee and may even contain supply chain malware attacks. 

Since scientific computing is Jupyter's main goal, security does not have the same priority. Loggings in Jupyter Notebooks mostly focus on application logs to track usability and/or performance issues instead of aiding in finding security vulnerabilities or anomalous, abusive user behaviors.  A community of volunteers is Jupyter's main security driver. For example, only six teen security-related bugs have been reported in Jupyter's Github, in which three bugs are still open~\cite{Issues·j31:online}. 

The Zeek project has started looking into parsing WebSocket (Pull request 3555~\cite{WebSocke51:online}); Companies such as NVIDIA/Amazon have started ~\cite{Howtoimp56:online,SecureSe68:online} building security assessment extension for Jupyter's Python extension; Government agencies such as NASA have published guidelines to secure Jupyter deployments~\cite{SecureSe68:online}. Design of the Instrumented SSH~\cite{campbell2011local} at NERSC and system provenance (Bates et al.) has the potential to guide our design of Jupyter kernel auditing. However, a holistic security approach is still missing~\cite{peiserttrustworthy}.

%% file: 08_conclusion.tex
\section{Conclusion}\label{conclusion}
Jupyter is the most popular data-intensive science application, and auditing its network security is critical. On balance, this is the first paper to systematically describe the threat model against Jupyter Notebooks and lay out the design of auditing Jupyter to have better visibility against such attacks. 

%% file: 99_ack.tex
\section{Acknowledgements}
The authors would like to thank NCSA staff for continuous support in data analysis and testbed deployment. This work utilized a security testbed at NCSA maintained by Phuong Cao and set up by Alexander Withers. We thank the TrustedCI leadership team, other TrustedCI fellows, notably Dr. Rick Wagner, Dr. Jim Basney, Anita Nikolich, and community members (ESnet, LBNL, and Zeek) for their valuable feedback and support in an early version of this manuscript. We also want to recognize the following people/organizations/programs: the NSF's Trusted CI Cybersecurity Center of Excellence, Illinois Computes, FABRIC, ACCESS, and Delta allocations, NCSA Integrated Cyberinfrastructure/IRST team, particularly Timothy Boerner, James Eyrich, and Ryan Walker.

This work was partly supported by the National Science Foundation (NSF) under contract \#2319190 and \#2430244. Any opinions, findings, conclusions, or recommendations expressed in this material are those of the authors and do not necessarily reflect the views of their employers or sponsors.